\documentclass{PoS}
\usepackage{amsmath,amssymb,bm}
\graphicspath{{./Figs/}}

\title{Path optimization method with use of neural network
for the sign problem in field theories\thanks{Report No.: YITP-18-120, KUNS-2738}}

\ShortTitle{Path optimization in field theories}

\author{\speaker{Akira Ohnishi}\\
	Yukawa Institute for Theoretical Physics, Kyoto University, Kyoto 606-8502, Japan\\
        E-mail: \email{ohnishi@yukawa.kyoto-u.ac.jp}}

\author{Yuto Mori\\
	Department of Physics, Faculty of Science, Kyoto University, Kyoto 606-8502, Japan\\
        E-mail: \email{mori.yuto.47z@st.kyoto-u.ac.jp}}

\author{Kouji Kashiwa\\
	Fukuoka Institute of Technology, Wajiro, Fukuoka 811-0295, Japan\\
        E-mail: \email{kashiwa@fit.ac.jp}}

\abstract{
We investigate the sign problem in field theories
by using the path optimization method with use of the neural network.
For theories with the sign problem,
integral in the complexified variable space is a promising approach
to obtain a finite (non-zero) average phase factor.
In the path optimization method,
the imaginary part of variables are given as functions of the real part,
$y_i=y_i(\{x\})$, and are optimized to enhance the average phase factor.
The feedforward neural network can be used to give and to optimize
functions with many variables.
The combined framework, the path optimization with use of the neural network,
is applied to the complex $\phi^4$ theory at finite density,
the 0+1 dimensional QCD at finite density,
and the Polyakov loop extended Nambu-Jona-Lasinio (PNJL) model,
all of which have the sign problem.
In these cases, the average phase factor is found to be enhanced significantly.
In the complex $\phi^4$ theory,
it is demonstrated that the number density is calculated at a high precision.
On the optimized path, the imaginary part is found to have strong
correlation with the real part on the temporal nearest neighbor site.
In the 0+1 dimensional QCD, we compare the results in two different treatments
of the link variable:
optimization after the diagonal gauge fixing
and optimization without the diagonal gauge fixing.
These two methods show consistent eigenvalue distribution
of the link variables.
In the PNJL model with homogeneous field ansatz, 
finite volume results approach the mean field results as expected,
and the phase transition behavior can be described.
}

\FullConference{The 36th Annual International Symposium on Lattice Field Theory - LATTICE2018\\
		22-28 July, 2018\\
		Michigan State University, East Lansing, Michigan, USA.}

\begin{document}

\section{Introduction}
When the action is complex,
strong cancellation occurs in integrating the Boltzmann weight at large volume.
This is referred to as the sign problem, and appears
in various problems in physics.
For example, let us consider the Fermion action at finite chemical potential.
The (relativistic) Fermion matrix $D$ has the $\gamma_5$ hermiticity,
$(\gamma_5 D(\mu) \gamma_5)^\dagger = D(-\mu^*)$,
then the Fermion determinant satisfies
$(\det D(\mu))^*=\det D(-\mu^*)$.
The Fermion determinant is real at zero chemical potential or
pure imaginary chemical potential, but is complex
at real finite chemical potential.
The Fermion determinant appears in the partition function,
and the effective action containing the Fermion determinant effects
is represented as $S_\mathrm{eff}=S_B - \log\det D(\mu) \in \mathbb{C}$,
where $S_B$ is the bosonic action.
Thus the integrand in the partition function with Fermions
is generally complex at finite density and becomes real only at zero density.

Finite density lattice QCD contains Fermions and 
the sign problem naturally appears~\cite{deForcrand}.
Then it is difficult to obtain precise predictions on
dense matter in atomic nuclei, neutron stars, their mergers and supernovae.
In heavy-ion collisions, baryon chemical potential is small
at LHC and the top energy of RHIC,
then the finite chemical potential effects
may be treated perturbatively
as given by the Taylor expansion from zero density.
By comparison, lower energy heavy-ion collisions are considered
to produce dense matter at finite baryon density.
In order to provide reliable information on dense matter
by the first principles method of QCD, the Monte-Carlo simulation
of lattice QCD,
perturbative treatment of chemical potential is not enough,
and we have to evade the sign problem.

There are many approaches to the sign problem
as discussed in the present lattice meeting:
Taylor expansion in $\mu/T$~\cite{Ratti,Sharma,Steinbrecher},
the imaginary chemical potential with analytical continuation
or for the canonical ensemble~\cite{Guenther,Goswami},
and the strong coupling approach~\cite{Unger,Klegrewe}
are the mature and useful methods to investigate finite density QCD,
but it seems that we cannot reach cold dense matter in the continuum limit.
Recently, integral methods in the complexified variable space
have been attracting attention:
the Lefschetz thimble method~\cite{Zambello},
the Complex Langevin method~\cite{Sinclair,Tsutsui,Attanasio,Ito,Joseph,Wosiek},
and the path optimization method~\cite{Lawrence,Warrington,Lamm}
are categorized to the complexified variable method.
These methods are still premature, but are developing rapidly.
It should be noted that other new approaches are also proposed
to tackle the sign problem~\cite{TsutsuiDoi,Ogilvie,Jaeger}.

In the present proceedings, we concentrate on the integration methods
in the complexified variable space.
Let us consider the complex Boltzmann weight $\exp(-S(x))$,
which is a holomorphic (complex analytic) function
of the complexified variable $x \to z$.
Then the Cauchy(-Poincare) theorem tells us that the partition function
is independent of the integral path
as long as the path is modified continuously without going through
the singular point.
One of the typical examples is the Gaussian integral,
which appears, for example, when we bosonize repulsive four-Fermi interaction
in the NJL model~\cite{Mori2018b},
\begin{align}
\int_\mathbb{R} d\omega e^{-\omega^2/2+i\rho_q\omega}
=
\int_{\mathbb{R}+i\rho_q} d\omega e^{-(\omega-i\rho_q)^2/2-\rho_q^2}
=\sqrt{2\pi}\exp(-\rho_q^2)
\ .
\end{align}
By shifting the integral path of the vector field ($\omega$)
in the imaginary direction,
a rapidly oscillating function at large density ($\rho_q$)
becomes a real function.

\begin{figure}
\centerline{\includegraphics[width=140mm,bb=20 360 765 575,clip]{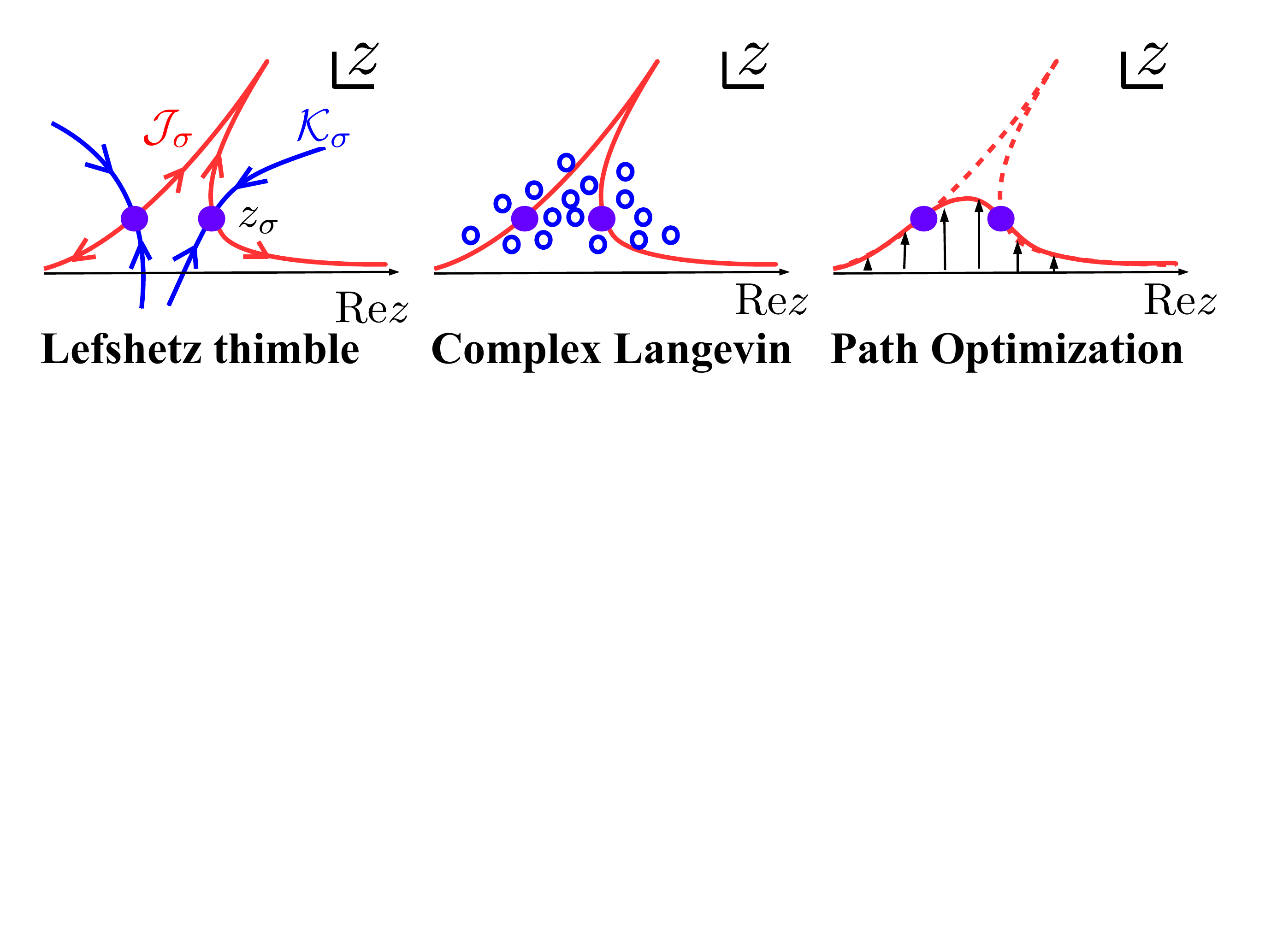}}
\caption{Schematic picture of integration path and sampled configurations
in the Lefschetz thimble method (left), the complex Langevin method (middle),
and the path optimization method (right).}
\label{Fig:path}
\end{figure}

As in the above case, if we can choose the integral path going through
the saddle point dominantly contributing to the integral,
we can avoid the integral of rapidly oscillating
function and we can avoid the sign problem.
This is a basic idea used in the method using the integral
over complexified variables.
In the Lefschetz thimble method (LTM)~\cite{Witten,Aurora,Fujii2013,GLTM},
the integral path (manifold) $J_\sigma$, referred to as the thimble(s),
is obtained by solving the flow equation,
$\dot{z}_i=\overline{\partial S/\partial z_i}$,
from the fixed point $\sigma$ ($\left.\partial S/\partial z_i\right|_\sigma=0$),
as shown in the left panel of Fig.~\ref{Fig:path}.
On the thimble, the imaginary part of the action is constant,
and the sign problem is weakened.
But we still have several problems in applying 
LTM
to field theories.
For example, the integral measure $d^Nz$ contains the complex phase
which is generally not constant (residual sign problem),
and when two or more thimbles contribute to the partition function,
the integrals on different thimbles can have different complex phases
and would cancel each other (global sign problem).
In addition, finding fixed points and constructing relevant thimbles
are not easy.
For the last point, a more practical method,
referred to as the generalized Lefschetz thimble method (GLTM),
to obtain relevant thimbles is proposed~\cite{GLTM}.
By 
evolving 
the integral path by using the flow equation
from the real axis (original integration path), the path becomes closer
to relevant thimbles. Even if we do not reach the thimble,
the phase fluctuation is generally suppressed on the evolved path.

The complex Langevin method (CLM)~\cite{Parisi1983,Klauder1983,Aarts2010,Nagata2016,Seiler2013,Ito2016,Sinclair,Tsutsui,%
Attanasio,Ito,Joseph} is another promising approach.
By solving the complex Langevin equation,
$\dot{z}_i=-\partial S/\partial z_i + \eta_i(t)$
with $\eta_i$ being the white noise,
we can generate configurations
around the fixed point as shown in the middle panel of Fig.~\ref{Fig:path},
and we can calculate observables as an ensemble average,
$\langle\mathcal{O}\rangle=\langle\mathcal{O}\rangle_\mathrm{CLM}$.
Since no phase reweighting is necessary,
there is no sign problem, in principle.
Thanks to recent developments of the gauge cooling technique~\cite{Seiler2013}
and the deformation technique to avoid the singular point
of the action~\cite{Ito2016}, 
one can perform stable simulations even at high density region.
One of the problems in the complex Langevin method
is the occasional convergence to wrong results.
When the magnitude distribution of the drift term has a long tail
than the exponential decay, the results are not reliable
even if they converge~\cite{Nagata2016}.
Unfortunately, the current CLM simulations are not reliable
in the phase transition region at low temperatures~\cite{Ito}.
We need further analyses to understanding the origin of the long tail.

In the last lattice meeting~\cite{Lat2017-AO} and in Ref.~\cite{Mori2017a},
we have proposed another method, the path optimization method (POM),
in which 
the integration path is optimized to evade the sign problem,
i.e. to enhance the average phase factor.
Since there is no singular point of the Boltzmann weight at finite $z$
in most of physical systems,
the integral of the Boltzmann weight is independent of the path.
(The exception is the lattice simulations with the Fermion determinant rooting,
$\sqrt{\det D}\,\exp(-S_G)$,
which gives rise to the cut in the Boltzmann weight.)
It should be noted that the zero point of the Fermion determinant
is the singular point of $S_\mathrm{eff}$,
but just a zero point of the Boltzmann weight $\exp(-S_\mathrm{eff})$.
Then the optimization can be done in various ways.
For example, using the flow equation is one of the ways
to optimize the path as in GLTM.
In one dimensional integral,
we can parameterize the imaginary part by a trial function of the real part,
and optimize the trial function
by the standard gradient descent method~\cite{Mori2017a}.
It is also possible to utilize the neural network~\cite{Mori2018a,Kashiwa2018a}.
Because of this flexibility, similar ideas have been applied to several
problems recently:
1+1 dimensional $\phi^4$ theory~\cite{Mori2018a},
0+1 dimensional $\phi^4$ theory~\cite{Bursa2018},
2+1 dimensional finite density Thirring model
(their method is referred to as the sign-optimized manifold method (SOMMe))%
~\cite{Lawrence,Warrington},
1+1 dimensional QED~\cite{Lamm},
and the Polyakov-loop extended Nambu-Jona-Lasinio (PNJL)
model~\cite{Kashiwa2018a}.

When we have many variables as in the field theories,
the neural network is a useful tool:
It can describe any functions of many variables,
then it is equivalent to prepare the complete set of trial functions.
In this proceedings, we apply the path optimization method
with use of the neural network
to field theories with the sign problem.
After a brief review of the path optimization method,
the neural network, and the stochastic gradient method,
we discuss
1+1 dimensional $\phi^4$ theory at finite $\mu$~\cite{Mori2018a},
0+1 dimensional QCD at finite $\mu$~\cite{Mori2018c},
and the Polyakov-loop extended Nambu-Jona-Lasinio (PNJL)~\cite{Kashiwa2018a}.

\section{Path optimization with use of neural network}

We consider the case where the Boltzmann weight $\exp(-S)$ is
a complex and analytic function of the field variables
$x=\{x_i; i=1, 2, \ldots N\}, x_i \in \mathbb{R}$
with $N$ being the number of variables.
By complexifying the field variables $z_i=x_i+iy_i$,
the Boltzmann weight becomes a holomorphic (complex analytic) function
of the complexified field variables $z$.
The partition function can be written as
\begin{align}
\mathcal{Z}
=\int_{\mathcal{C}_\mathbb{R}} d^Nx\, \exp(-S(x))
=\int_{\mathcal{C}_\mathbb{C}} d^Nz\, \exp(-S(z))
=\int_{\mathcal{C}_\mathbb{R}} d^Nx\, J(z) \exp(-S(z))
\ ,
\end{align}
where $J(z)=\det(\partial z_i/\partial x_j)=\det(\delta_{ij}+i\partial y_i/\partial x_j)$ is a Jacobian.
We here assume that the imaginary part of the complexified variables are
given as functions of the real part, $y_i=y_i(x)$,
which specify the integration path.
If we can optimize the functions $y_i(x)$ to enhance the average phase factor
to be clearly above zero, precise calculation of observables becomes possible.

In the path optimization method,
we optimize the integration path specified by $y(x)=\{y_i(x); i=1, 2, \ldots N\}$
to minimize the cost function, which represents the seriousness of the sign problem.
We adopt the following cost function,
\begin{align}
\mathcal{F}[y]= \mathcal{Z}_\mathrm{pq}-|\mathcal{Z}|
= |\mathcal{Z}|\left(|\langle{e^{i\theta}}\rangle|^{-1}-1\right)
\ ,
\end{align}
where $\theta$ is the complex phase of the statistical weight,
$W=J(z)\exp(-S(z))$,
and $\mathcal{Z}_\mathrm{pq}$ is the phase quenched partition function.
Since the partition function is independent of the path,
the above cost function is a monotonically decreasing function 
of the average phase factor.

Optimization of the path can be performed
via the gradient descent method or by using the neural network
which is used in machine learning.
In the one-dimensional integral case,
we can expand the imaginary part by a complete set of functions,
$y(x)=\sum_n c_n H_n(x)$ with $\{H_n\}$ being a complete set,
and tune the coefficients $c_n$ to minimize the cost function
using the gradient descent method.
In Ref.~\cite{Mori2017a},
we have found that the optimized path is close to the thimble
around the fixed points.
Thus the path optimization may be also regarded as a practical method
to search for the thimble.
On the optimized path, the rapid oscillation of the integrand
is suppressed, while the absolute values becomes smaller.
The hybrid Monte Carlo sampling works well on the optimized path,
and it is possible to calculate the observables precisely.

When we have many integral variables as in the field theories, 
it is tedious and practically impossible
to prepare the complete set of functions and to optimize the path.
The number of expansion parameters is $M^N$,
where $M$ is the number of functions for each degrees of freedom
needed for conversion.
We need to avoid the exponential growth of the number of parameters.

\begin{figure}
\centerline{
\includegraphics[width=60mm,bb=150 140 475 485,clip]{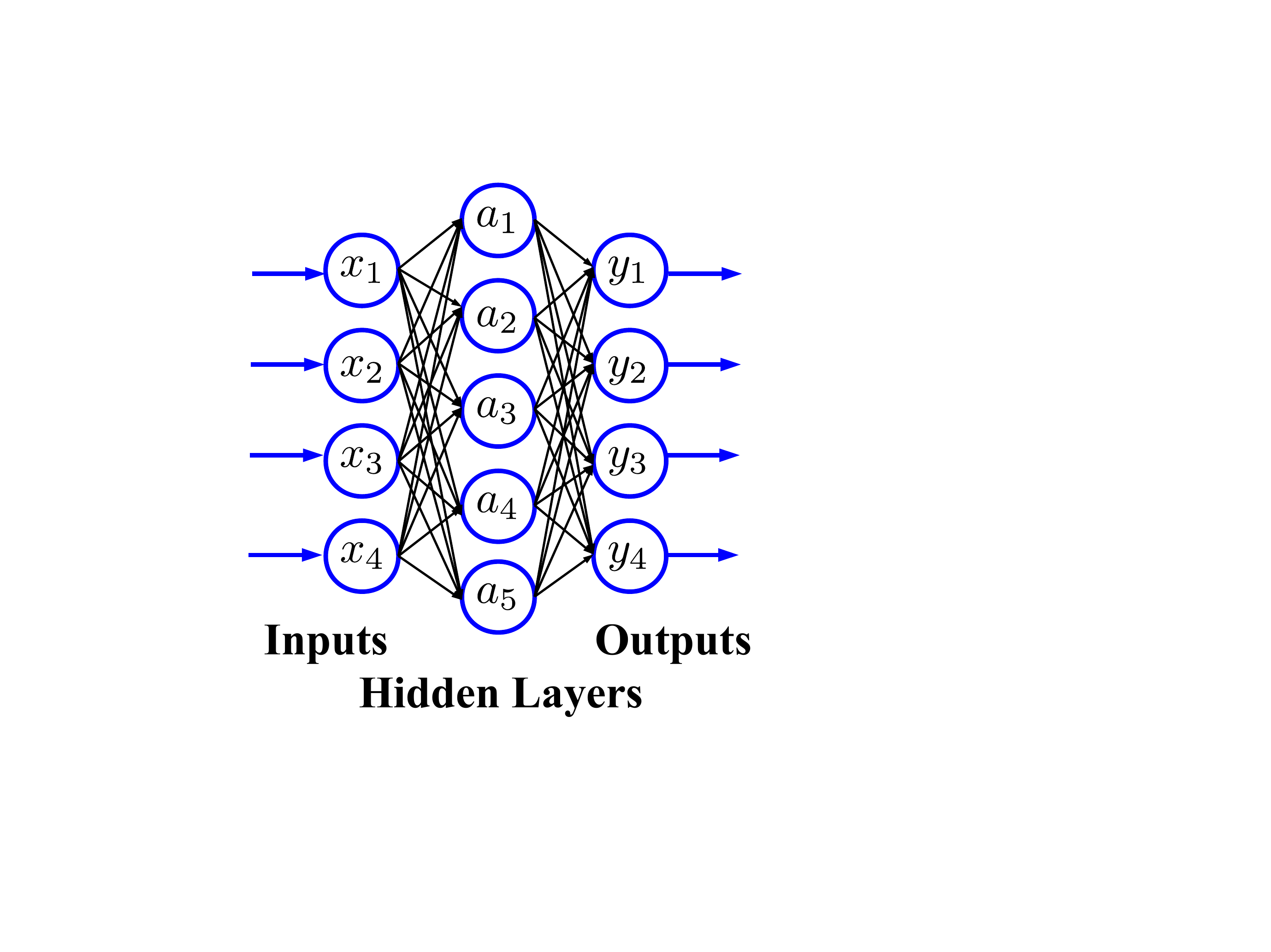}
}
\caption{Schematic picture of the neural network.}
\label{Fig:neural}
\end{figure}

One of the ways to avoid the above exponential growth
is to utilize the feedforward neural network.
The neural network is a mathematical model of the human brain.
It contains the input, hidden, and output layers,
and each layer consists of units which mimic neurons
as schematically shown in Fig.~\ref{Fig:neural}.
Units in one layer is connected with those in the previous and next layers,
and the variables are modified by combining the linear and non-linear
transformations.
In the single hidden layer case, these transformations are given as
\begin{align}
&a_i = g(W^{(1)}_{ij}x_{j} + b^{(1)}_i)\ , 
\ \ 
y_i  = \alpha_i (g(W^{(2)}_{ij}a_{j} + b^{(2)}_i)) + \beta_i\ , 
\end{align}
where $g(x)$ is called the activation function
and we adopt $g(x)=\tanh x$ in order to keep the differentiability.
We input $\{x_i\}$ and obtain $\{y_i\}$ as the outputs.
$W^{(1,2)}_{ij}$, $b^{(1,2)}_i$, $\alpha_i$ and $\beta_i$
are the parameters in the neural network.
The number of parameters is $\mathcal{O}(N^2)$
where $N$ is the number of variables,
provided that the number of units in the hidden layer is
proportional to $N$ 
and the number of layers is $\mathcal{O}(1)$.
The universal approximation theorem~\cite{Universal}
tells us that any functions can be described
in the large number limit of the units in the hidden layers,
even in the case of a single hidden layer.

In updating the parameters in the neural network,
we apply the stochastic gradient method.
We adopt the ADADELTA algorithm~\cite{ADADELTA},
\begin{align}
F_i^{(j)}&= \partial \mathcal{F}/ \partial c_i^{(j)}\ ,\\
r_i^{(j+1)} &= \gamma r_i^{(j)} + (1-\gamma) (F_i^{(j)})^2\ , \\
v_i^{(j+1)} 
&= \frac{\sqrt{s_i^{(j)}+\epsilon}}{\sqrt{r_i^{(j+1)}+\epsilon}} F_i^{(j)}\ , \\
c_i^{(j+1)} &= c_i^{(j)} - \eta v_i^{(j+1)}\ ,\\
s_i^{(j+1)} &= \gamma s_i^{(j)} + (1-\gamma) (v_i^{(j+1)})^2\ , 
\end{align}
where $c_i^{(j)}$ represents the $i$-th parameter in the $j$-th step of updates,
and $\mathcal{F}$ is the cost function.
Gradients $F_i^{(j)}$ are evaluated as the average
in a small number ($N_\mathrm{batch}$) of configurations (mini-batch training),
and their squared average is stored in $r_i^{(j)}$.
The parameters are updated by using the "velocity" $v_i^{(j)}$,
gradients normalized by the root mean square in the history of updates.
The squared averages of $v_i^{(j)}$ are stored in $s_i^{(j)}$.
The learning rate $\eta$ plays the role of the time step,
and $\gamma$ is referred to as the decay rate in machine learning
but is the survival rate in the actual meaning.
When the derivative and velocity become smaller
than the cutoff parameter $\epsilon$ during many updates,
the ADADELTA algorithm becomes 
equivalent to
the standard gradient descent
method, $\dot{c}_i=-\partial \mathcal{F}/\partial c_i$.

Usually, the neural network is trained by using the teacher data,
where the answers are prepared (supervised learning).
By comparison, we do not know the optimized path in advance,
then the parameters in the neural network are updated
only by the gradient of the cost function.
This update corresponds to the so-called unsupervised learning.

We generate the Monte Carlo configurations by the hybrid Monte Carlo (HMC).
We regard the real part of variables as the dynamical variable,
where the molecular dynamics Hamiltonian is
$H(x,p)=p^2/2+\mathrm{Re} S(z(x))$.
The derivative of the Jacobian requires numerical cost and is ignored
in the molecular dynamics evolution.
We solve the canonical equation of motion,
$\dot{x}=p$ and $\dot{p}=-\partial H/\partial x$,
and make Metropolis judgement with the Jacobian effects included.
We prepare $N_\mathrm{config}$ configurations,
update parameters using $N_\mathrm{batch}$ configurations
in each mini-batch training.
After updating parameters $N_\mathrm{config}/N_\mathrm{batch}$ times,
we generate $N_\mathrm{config}$ configurations
using the updated neural network parameters.
We repeat this update process
$N_\mathrm{epoch}$ times.
One epoch means the number of updates
where the $N_\mathrm{config}$ configurations are used
in the mini-batch training.

The path optimization with use of the neural network
has been applied to the one-dimensional integral~\cite{Lat2017-AO}.
The obtained optimized path is close to the thimble
and the path optimized by using the standard gradient descent method
around the fixed points.
These paths do not agree off the fixed points.
The statistical weights are small far off the fixed points,
and do not contribute to the observable calculation much.

\section{Application to field theory: complex $\phi^4$ theory}

Now let us apply the path optimization method with use of the neural network
to the 1+1 dimensional complex $\phi^4$ theory at finite $\mu$.
The Lagrangian of the complex $\phi^4$ theory is given as
$\mathcal{L}=\partial_\mu \phi^* \partial^\mu \phi - m^2 \phi^* \phi
- \lambda(\phi^*\phi)^2$,
and the action at finite $\mu$ on the Euclidean lattice is given as
\begin{align}
S = \sum_x & \left [ \frac{(4+m^2)}{2}\phi_{a,x}\phi_{a,x}
 + \frac{\lambda}{4}(\phi_{a,x}\phi_{a,x})^2
 - \phi_{a,x}\phi_{a,x+\hat{1}} 
 - \cosh\mu\, \phi_{a,x}\phi_{a,x+\hat{0}}
+i \epsilon_{ab} \sinh\mu\, \phi_{a,x}\phi_{b,x+\hat{0}}
 \right]
\ ,\label{Eq:phi4}
\end{align}
where $\phi=(\phi_1 + i \phi_2)/\sqrt{2}$
with $\phi_a (a=1,2)$ being the real scalar field.
The last term of the action gives rise to the imaginary part,
and we have the sign problem.
We complexify the real and imaginary parts,
$\phi_{1,x}=\varphi_{1,x}+i\xi_{1,x}$
and $\phi_{2,x}=\varphi_{2,x}+i\xi_{2,x}$,
respectively, and apply the path optimization method.

\begin{figure}
\centerline{
\includegraphics[width=70mm,bb=0 0 358 252,clip]{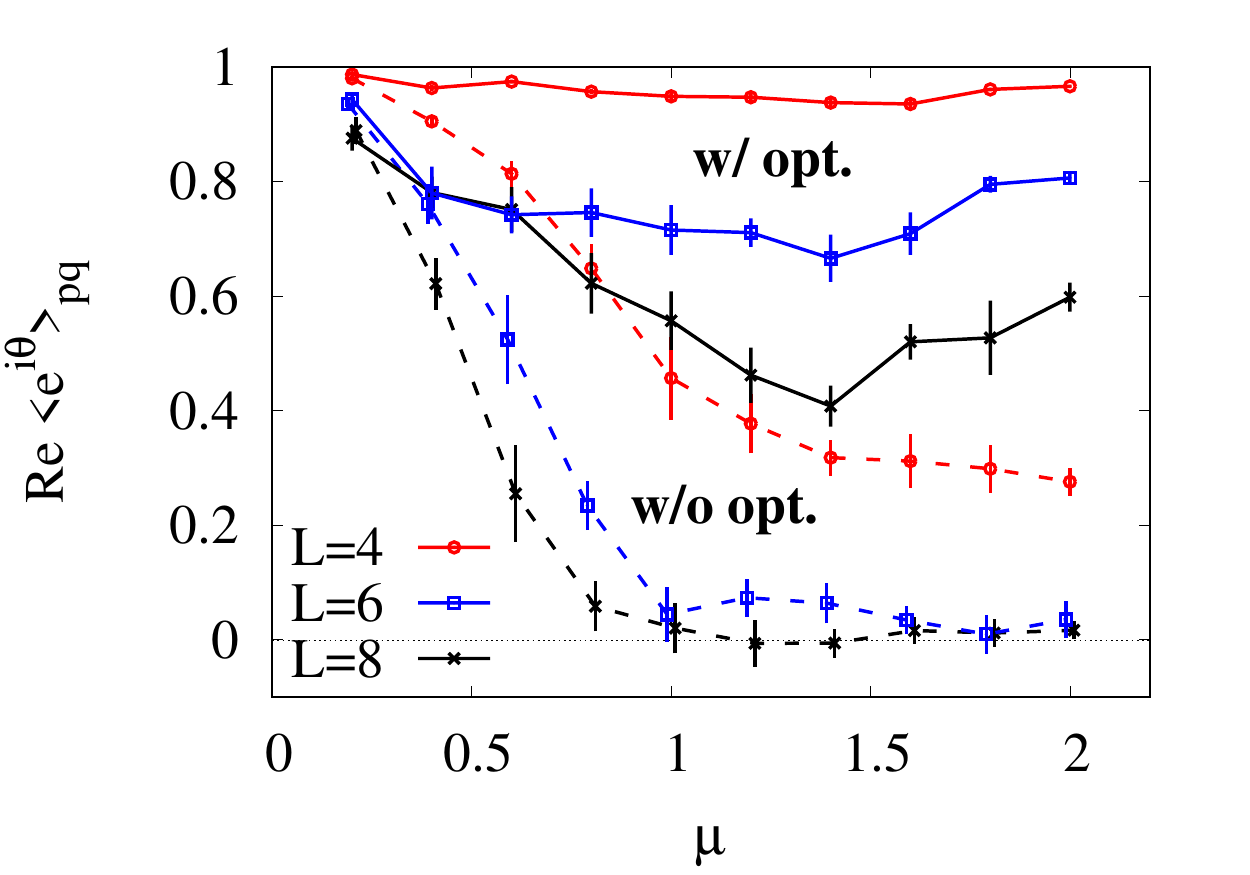}
\includegraphics[width=70mm,bb=0 0 360 250,clip]{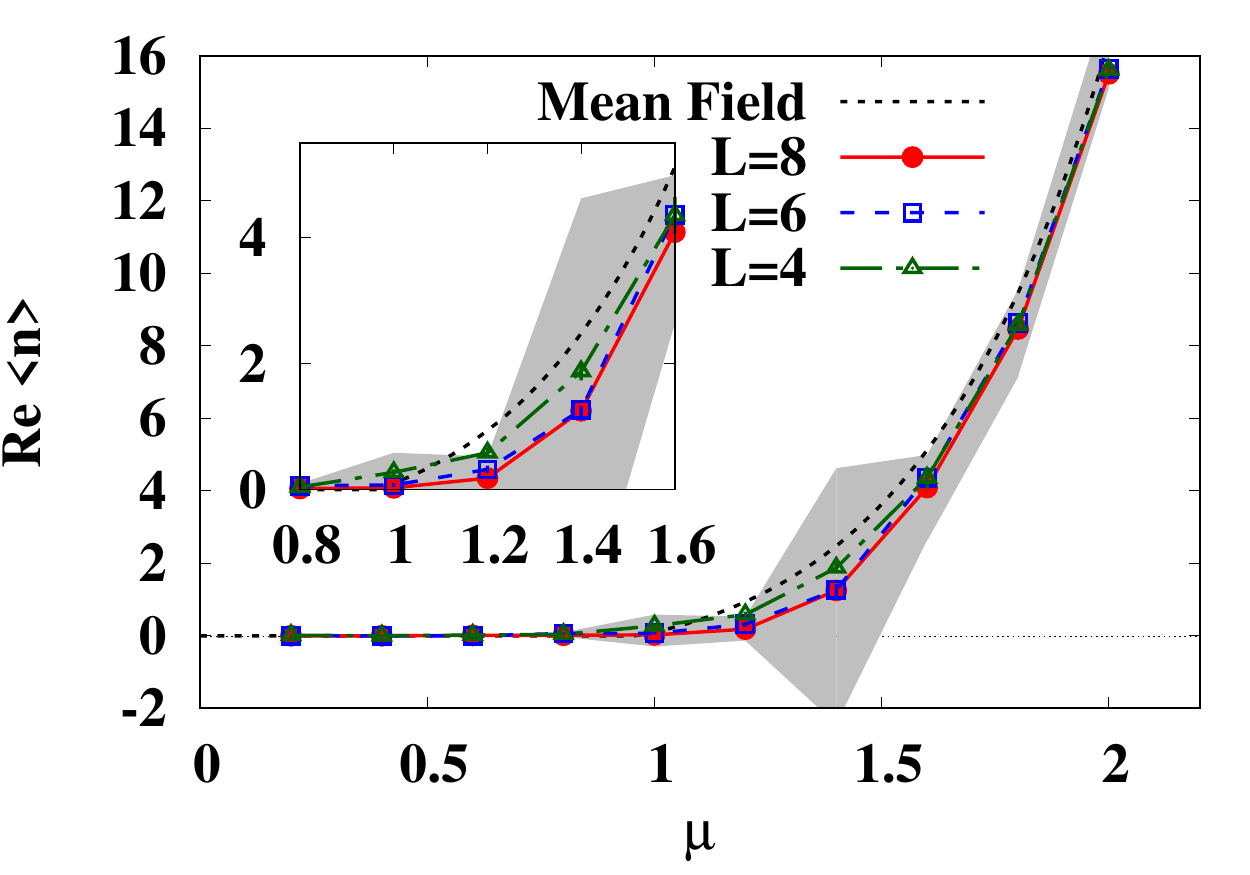}
}
\caption{Average phase factor (left) and number density (right)
in the complex $\phi^4$ theory at finite $\mu$.}
\label{Fig:phi4}
\end{figure}


We have performed the optimization of the integration path
on $4^2, 6^2$ and $8^2$ lattices
with $m=\lambda=1$
by using the
neural network with a single hidden layer. 
In the left panel of Fig.~\ref{Fig:phi4}, we show the average phase factor
as a function of $\mu$.
%
Without
the optimization, the average phase factor quickly decreases
towards zero.
By comparison, the HMC simulation on the optimized path shows that
the average phase factor is enhanced and kept to be greater than 0.4.
Then the statistical error of the expectation value
for a given number of configurations is much smaller on the optimized path
(lines) than that on the original path (gray area)
as shown in the right panel of Fig.~\ref{Fig:phi4}.
The calculated number density, $n=-\partial(S/V)/\partial \mu$,
starts to grow at around $\mu=1$, as expected from the mean field results,
$\mu_c(\mathrm{MF})=\mathrm{arccosh}\,(1+m^2/2)\simeq 0.96$,
while the growth is delayed compared with the mean field results
especially on the larger lattices.

These features qualitatively agree with the results
in previous works~\cite{Fujii2013,Aarts2009},
where 3+1 dimensional complex $\phi^4$ theory is discussed
using CLM~\cite{Aarts2009} and LTM~\cite{Fujii2013}.
It is found that the rapid decay of the average phase factor 
on the original path or in the phase quenched simulations,
enhanced average phase factor on the thimble~\cite{Fujii2013},
and delay of the density growth due to the interaction.

\begin{figure}
\centerline{\includegraphics[width=70mm,bb=105 75 660 490,clip]{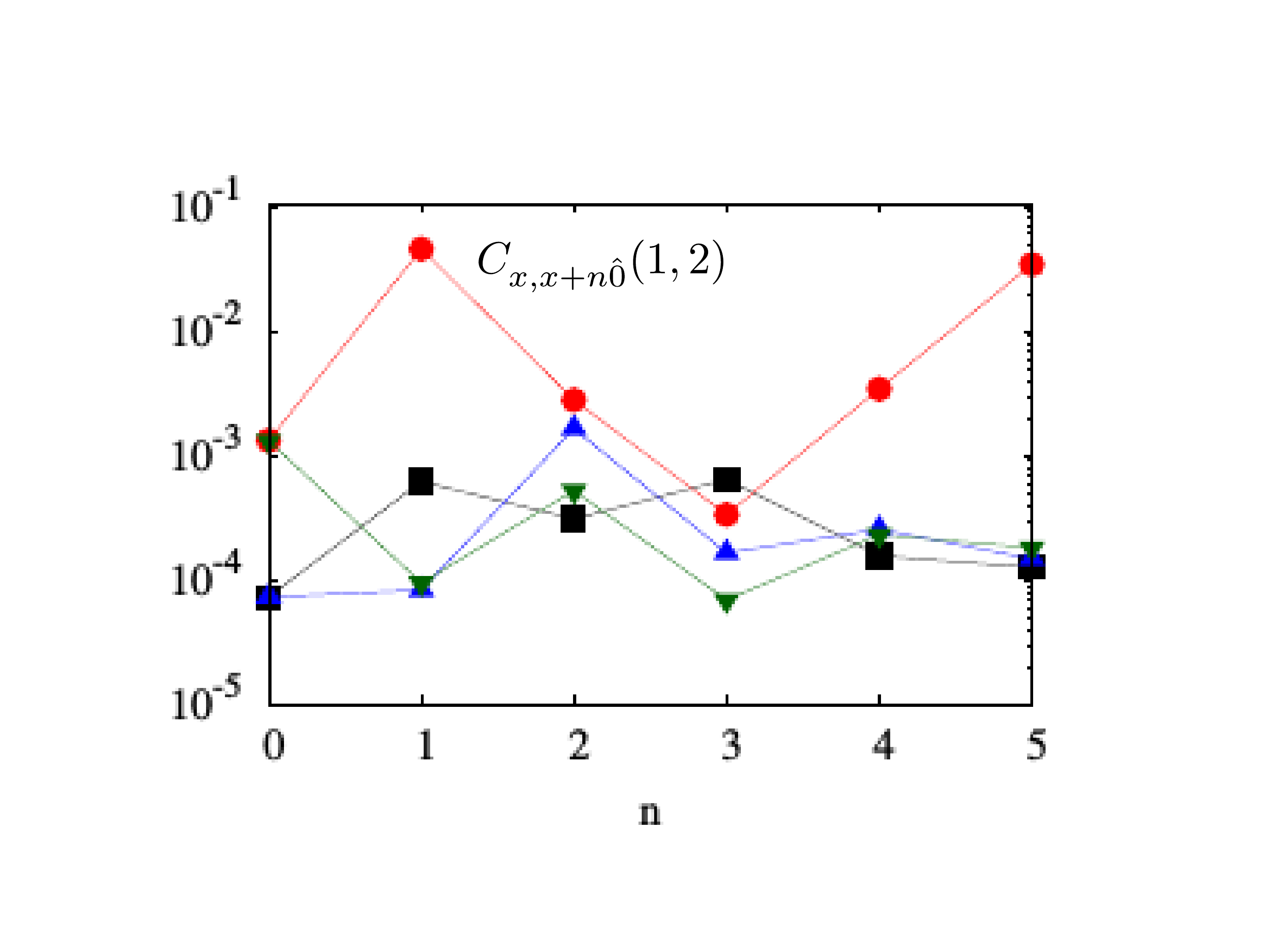}}
\caption{Correlation between the imaginary and real parts, $C_{xy}(a,b)$,
on a $6^2$ lattice.}
\label{Fig:corr}
\end{figure}

The average phase factor may not be large enough
compared with the LTM results~\cite{Fujii2013}.
This should be due to incomplete optimization.
Good ansatz helps us to enhance the average phase factor. 
For example, Bursa and Kroyter have proposed several ansatz
in the 0+1 dimensional complex $\phi^4$ theory
using the translational invariance
and the $\mathrm{U}(1)$ symmetry~\cite{Bursa2018}.
By using their ansatz, the average phase factor is well above 0.9 at $L=8$
with $L$ being the temporal lattice size.
The ansatz in Ref.~\cite{Bursa2018} assumes that the imaginary part
of the variable on a site is a function of the real part on the same 
and the nearest neighbor sites.
Also in our calculation, the correlation
$C_{xy}(a,b)
=(\partial \xi_{a,x}/\partial \varphi_{b,y})^2
+(\partial \xi_{b,y}/\partial \varphi_{a,x})^2$
is found to be strong for the nearest neighbor sites in the temporal direction,
$x$ and $x+\hat{0}$, as shown in Fig.~\ref{Fig:corr},
and other correlations are smaller by about 2 orders of magnitude.
Thus assuming the dependence
$\xi_{a,x}=f_a(\varphi_x,\varphi_{x\pm\hat{0}})$
would be a good starting point of unsupervised learning.

\section{Application to gauge theory: 0+1 dimensional QCD}

As the first step of application to gauge theory,
we discuss 0+1 dimensional QCD
with one species of staggered Fermion at finite $\mu$ on a $1\times N_\tau$
lattice%
~\cite{Bilic1988,Ravagli2007,Aarts2010b,Bloch2013,Schmidt2016,DiRenzo2017}.
The lattice action is given as
\begin{align}
S=&\frac12\! \sum_\tau \left(
	 \bar{\chi}_\tau e^\mu U_{\tau}\chi_{\tau+\hat{0}}\!
	-\!\bar{\chi}_{\tau+\hat{0}} e^{-\mu} U^{-1}_{\tau}\chi_\tau
	\right)
	\!+\!m\sum_\tau\!\bar{\chi}_\tau \chi_\tau
=\frac12 \bar{\chi}D\chi
\ .
\end{align}
The partition function is obtained as
\begin{align}
\mathcal{Z}=& \int \mathcal{D}U \det{D[U]}
=\int dU \det\left[X_{N_\tau}\!+\!(-1)^{N_\tau} e^{\mu/T} U
+\!e^{-\mu/T} U^{-1}
\right]
\nonumber\\
=&\frac{\sinh[(N_c+1)E/T]}{\sinh[E/T]}+2\cosh(N_c\mu/T)
\ ,
\\
X_{N_\tau}=&2\cosh(E/T)
\ ,\ 
E=\mathrm{arcsinh}\,m
\ ,\ 
U=U_1U_2\cdots U_{N_\tau}
\ ,\ 
T=1/N_\tau
\ .
\end{align}
It should be noted that only the product of link variables,
$U=U_1U_2\cdots U_{N_\tau}$, remains in the partition function,
then 0+1 dimensional QCD reduces to a one link problem.

The 0+1 dimensional QCD is a toy model,
but it contains the temporal hopping term which is the actual source
of the sign problem in 3+1 dimensional QCD.
The properties of the above partition function is also studied well
in the context of the strong coupling lattice QCD%
~\cite{SCLQCD,deForcrand2014}.

By using the residual gauge degrees of freedom,
it is also possible to take the diagonal gauge.
The link variable for color $\mathrm{SU}(3)$ in the diagonal gauge
is given as
$U = \mathrm{diag} (e^{ix_1},e^{ix_2},e^{ix_3})$ with $x_1+x_2+x_3=0$.
After complexification of gauge variables,
the partition function is now given as
\begin{align}
\mathcal{Z}=&
\int dU e^{-S}
= \int dx_1 dx_2 J H e^{-S}
\nonumber\\
=& \int dx_1 dx_2 
\left[\det \left(\frac{\partial z_a}{\partial x_b} \right) \right]
\left[\frac{8}{3\pi^2}\prod_{a<b}\sin^2\left(\frac{z_a-z_b}{2}\right)\right]
\left[\prod_a\left(X_{N_\tau} + 2\cos(z_a -i \mu)\right)\right]
\ ,
\end{align}
where the color index ($a,b$) runs from 1 to $N_c(=3)$,
and $z_1=x_1+iy_1, z_2=x_2+iy_2$ and $z_3=-(z_1+z_2)$.
The part in the first square bracket represents the Jacobian $J$,
the second is the Haar measure $H$,
and the third is the Boltzmann weight $\exp(-S)=\det D$.
We have assumed that $N_\tau$ is an even number.

We have only two variables in the diagonal gauge, 
then it is possible to work on the two dimensional mesh points
of $(x_1,x_2)$.
We first discuss the results where
the imaginary parts $(y_1,y_2)$ themselves are treated as the parameters.
The gradient descent equation is then given as
\begin{align}
\frac{dy_i(x_1,x_2)}{dt}=-\frac{\partial \mathcal{F}}{\partial y_i(x_1,x_2)}\ ,
\end{align}
where $t$ is the fictitious time.

\begin{figure}
\centerline{\includegraphics[width=70mm,bb=10 10 350 245,clip]{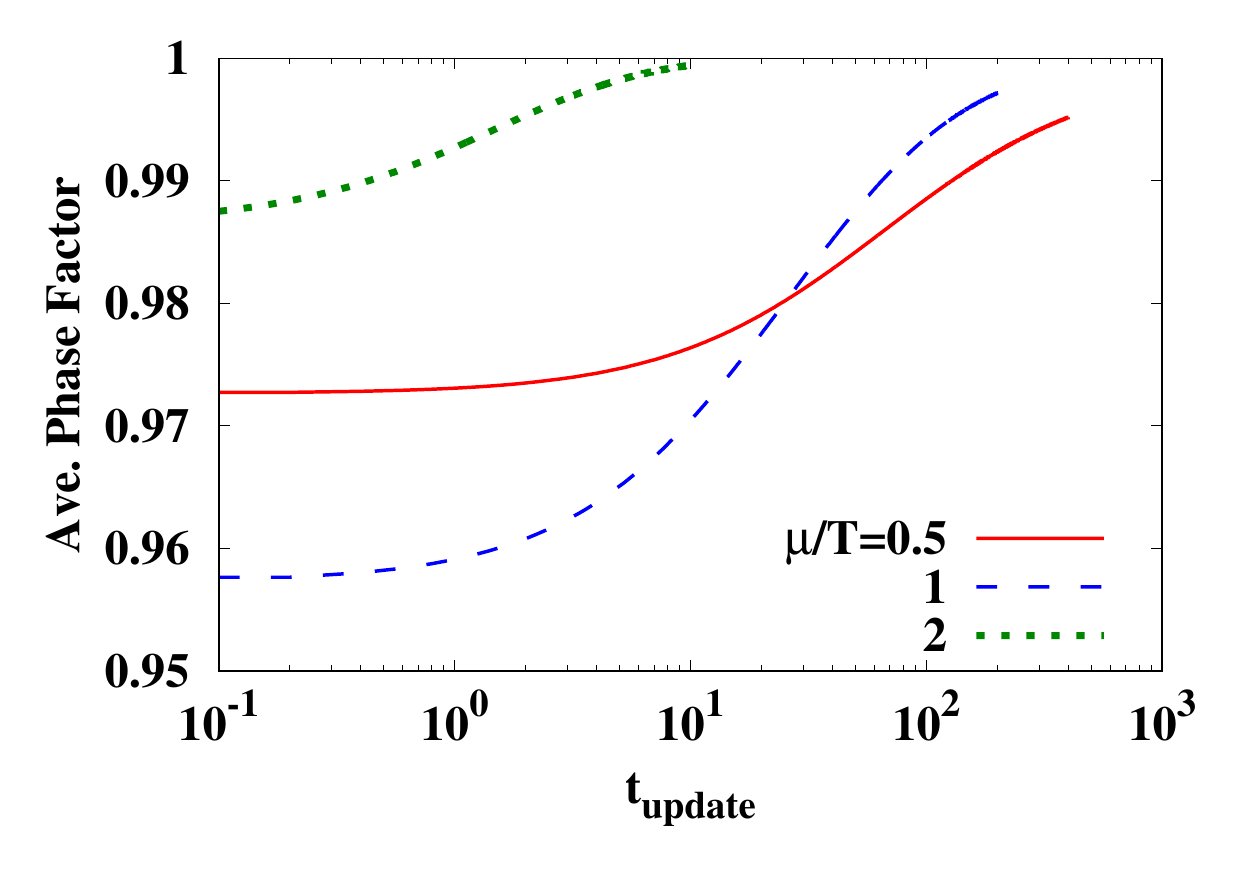}}
\caption{Average phase factor in 0+1 dimensional QCD.}
\label{Fig:QCD01APF}
\end{figure}

\begin{figure}
\centerline{\includegraphics[width=120mm,bb=0 0 310 230,clip]{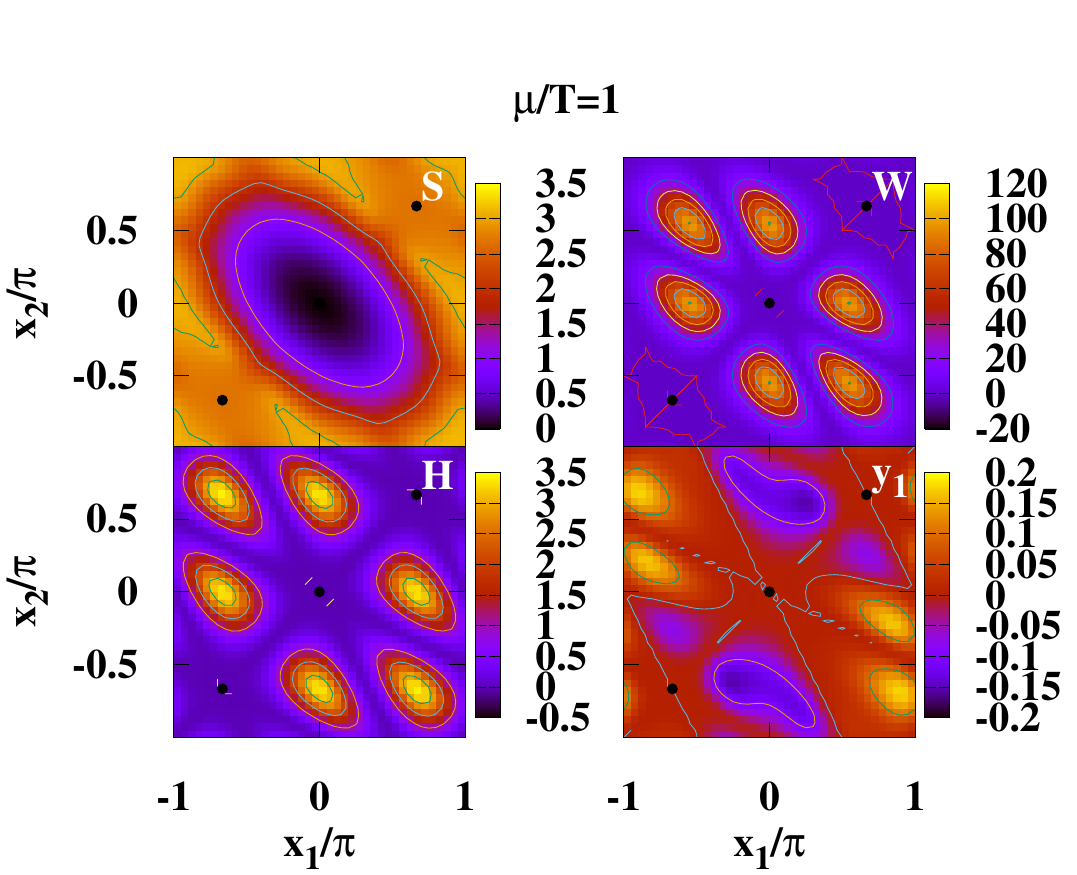}}
\caption{Action $S$ (from the minimum, left top),
Haar measure $H$ (left bottom), the statistical weight $He^{-S}$,
and the imaginary part of variable $y_1$
in 0+1 dimensional QCD.}
\label{Fig:QCD01dist}
\end{figure}

We have performed path optimization for 0+1 dimensional QCD
with $T=1/2$ and $m=0.05$.
By optimizing the path, the average phase factor goes above 0.99 easily
as shown in Fig.~\ref{Fig:QCD01APF}.
It is partly because the average phase factor on the original path
is above 0.95. Nevertheless we would like to emphasize
that the reduction of $(1-\mathrm{APF})$ is significant.
In the 3+1 dimensional calculation, the 0+1 dimensional QCD action
appears in each spatial point on the lattice,
then the average phase factor in 3+1 dimensions on a $L^3\times N_\tau$ lattice
would be around the $L^3$-th power of that in the 0+1 dimensions,
$\mathrm{APF}_{3+1}\simeq (\mathrm{APF}_{0+1})^{L^3}$,
provided that other action terms do not make it worse.
Then $\mathrm{APF}_{0+1}=0.95$ leads
to $\mathrm{APF}_{3+1}\simeq 4\times 10^{-12}$ on a $8^3\times N_\tau$ lattice,
while $\mathrm{APF}_{0+1}=0.995$
gives $\mathrm{APF}_{3+1}\simeq 0.08$.
This average phase factor is not large but it would be enough to obtain
some meaningful results.

The results of path optimization for $\mu/T=1$
is shown in the right-bottom panel of Fig.~\ref{Fig:QCD01dist}.
The imaginary part $y_1$ is modified from zero (original path)
to be in the range of $-0.2 < y_1 < 0.2$ after optimization.
($y_2$ is obtained by exchanging $x_1$ and $x_2$ axes.)
The amount of the shift is roughly proportional
to $-\partial \mathcal{F}/\partial y_i$, as expected from perturbation.
The action shows a deep minimum at $(x_1,x_2)=(0,0)$,
and there are two local minima at $(x_1,x_2)=\pm(2\pi/3,2\pi/3)$.
By comparison, the Haar measure divides the plain
into the six separated regions.
As a result, the probability distribution, the real part of $W=H\exp(-S(z))$
(except for the Jacobian), also has six regions.

The diagonal gauge is useful as shown above,
but there are two problems.
One is that it is not always possible to take this gauge for all links.
When we have spatial dimensions, we can take the diagonal gauge
for temporal links but not for spatial links.
The other is the separated probability distribution.
The separated distribution does not cause trouble
in mesh point integral, but is problematic in HMC simulations.
Since it is difficult to overcome the statistical barrier from the Haar measure,
we need to invoke the exchange Monte Carlo or different 
tempering~\cite{Fukuma2017,AlexandruTempering}
in order to sample configurations in the above six regions equally well.

\begin{figure}
\centerline{\includegraphics[width=60mm,bb=50 5 255 220,clip]{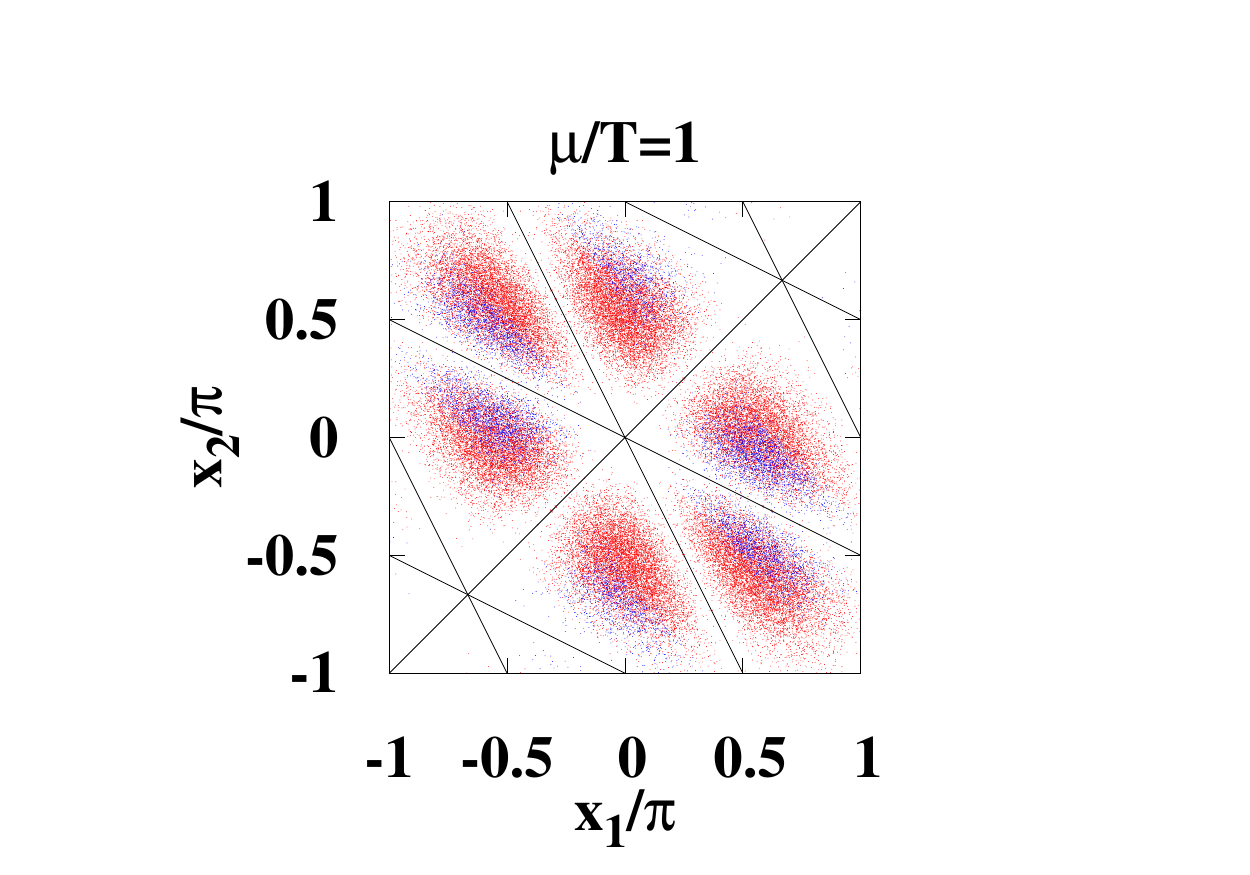}}
\caption{Distribution of $(x_a,x_b)$ ($a\not=b$) in HMC simulations
in 0+1 dimensional QCD.}
\label{Fig:QCD01MCdist}
\end{figure}

Now we proceed to link variable sampling without diagonal gauge fixing.
The $\mathrm{SU}(3)$ link variable $U$ is complexified 
to a $\mathrm{SL}(3)$ variable $\mathcal{U}$.
We here adopt the following complexification,
\begin{align}
U \in \mathrm{SU}(3)
\to \mathcal{U}(U)=U\,\prod_{a=1}^{N_c^2-1} e^{y_a\lambda_a/2}
\in \mathrm{SL}(3)
\ .
\end{align}
This form of the link variable is convenient to calculate
the derivative of the matrix element with respect to the real part of variables
as well as the imaginary part.
Since we have eight variables,
mesh point integral is not possible.
We adopt HMC for sampling, and the neural network for optimization.
In HMC, we regard the $\mathrm{SU}(3)$ part is regarded as the dynamical
variable of the molecular dynamics, and the Hamiltonian is taken to be
$H=P^2/2 + \mathrm{Re} S(\mathcal{U}(U))$, where $P$ is the conjugate 
momentum matrix of $U$.

We have performed the path optimization using a neural network
for the link variable without gauge fixing~\cite{Mori2018c}.
After optimization,
we generate configurations in HMC using the optimized neural network.
The $\mathrm{SL}(3)$ link variables can be diagonalized
by the similarity transformation, $\mathcal{U} \to P^{-1}\mathcal{U}P$.
The diagonalized link variables are given as
$\mathrm{diag}(e^{iz_1},e^{iz_2},e^{iz_3})$ with $z_1+z_2+z_3=0$.
In Fig.~\ref{Fig:QCD01MCdist}, we show the distribution
of $(x_a,x_b)$ ($a\not= b$) with $x_a=\mathrm{Re}\,z_a (a=1,2,3)$.
We show the distribution with $(a,b)=(1,2)$ with blue dots
and other combinations of $(a,b)$ with red dots.
This distribution (blue+red) is consistent with that
in the probability distribution in the diagonalized gauge.
While the order of the eigenvalues may be easily exchanged
in the diagonalization, 
we can deduce that the six separated regions are visited in the HMC sampling.
This is natural because the six regions are related with each other
by the exchange $z_a \leftrightarrow z_b (a,b=1,2,3)$
and by the symmetry $S(-z)=(S(z^*))^*$ which corresponds
to the transformation of $x_a \to -x_a$,
and the the eight variable link variable contains these transformations.

We have also confirmed that we can reproduce the exact results of
the chiral condensate, quark number density, and the Polyakov loop
in both of the treatment of link variables within the statistical errors.
These results will be reported in the forthcoming paper~\cite{Mori2018c}.

\section{Application to QCD effective model with phase transition:
PNJL model with homogeneous ansatz}

One important task in tackling the sign problem
is how to handle the multi thimble contributions.
When we have two or more fixed points contributing to the partition function
significantly, jumping from the one to other in HMC simulations
requires more elaborate procedures.
The separated regions of probability in the 0+1 dimensional QCD
are apparent ones and are connected in the $\mathrm{SL}(3)$ link variables.
By comparison, one expects the 
phase transition in dense QCD,
where two different configurations of the chiral condensate
and the Polyakov loop, hadronic and quark matter, compete
at around the phase transition boundary.
This would correspond to the two fixed point problem.
In other words, two thimbles starting from these
two fixed points would contribute significantly to the partition function.

As an example to discuss the phase transition,
the Polyakov loop extended Nambu-Jona-Lasinio (PNJL) model~\cite{Fukushima}
seems to be nice.
In the mean field approximation, we find the first order phase transition
boundary in low $T$ and finite $\mu$ region.
When we take account of fluctuations of fields,
the PNJL model has the sign problem.
We have applied the path optimization method to the PNJL~\cite{Kashiwa2018a}.
The Fourier transformed field variables are truncated at zero momentum,
i.e. homogeneous field ansatz is used,
while the volume is assumed to be finite.
In this setup, the finite volume results should converge
to the mean field results in the large volume limit.

\begin{figure}
\includegraphics[width=70mm,bb=0 0 480 462,clip]{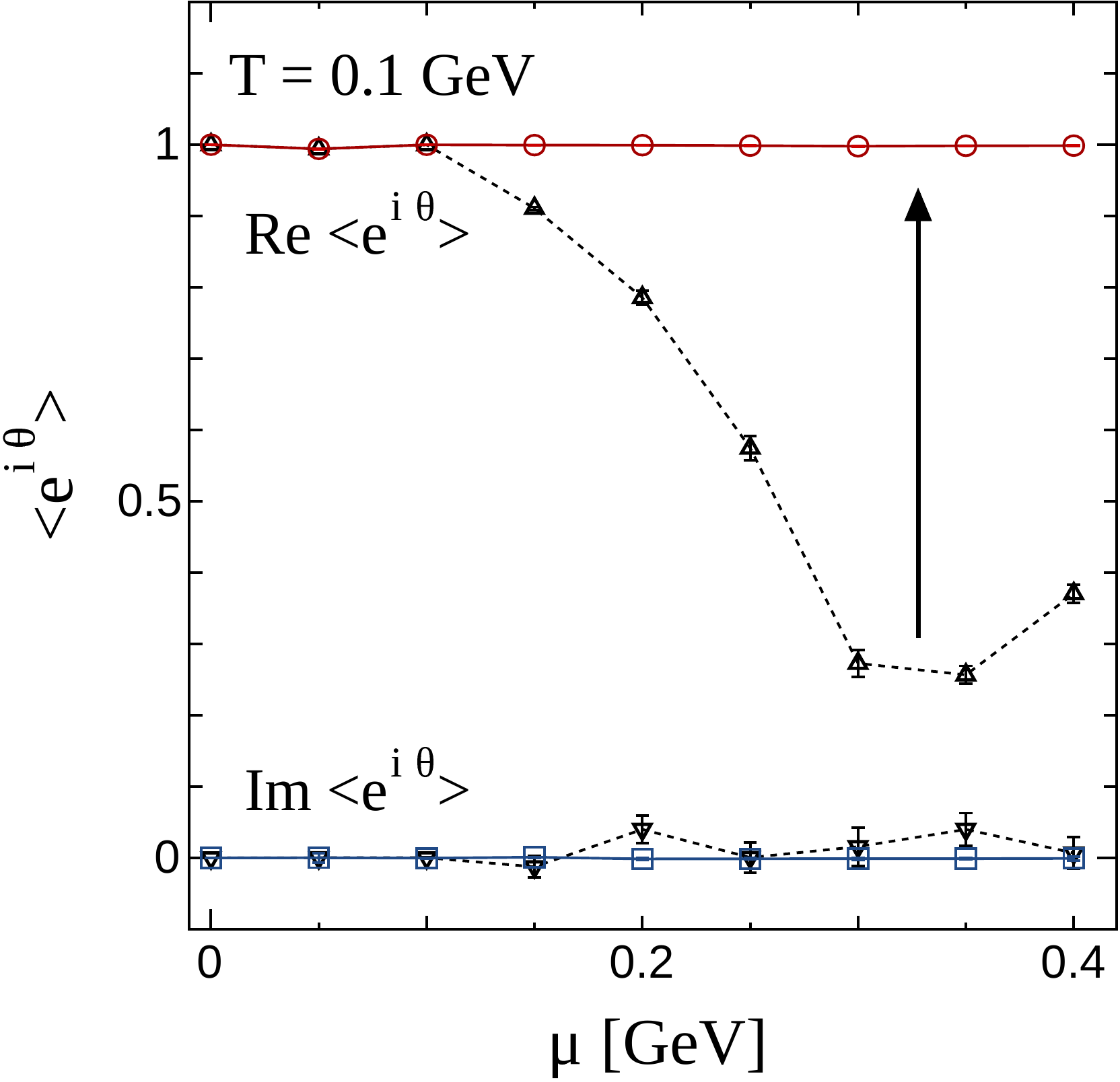}
\includegraphics[width=70mm,bb=0 0 485 470,clip]{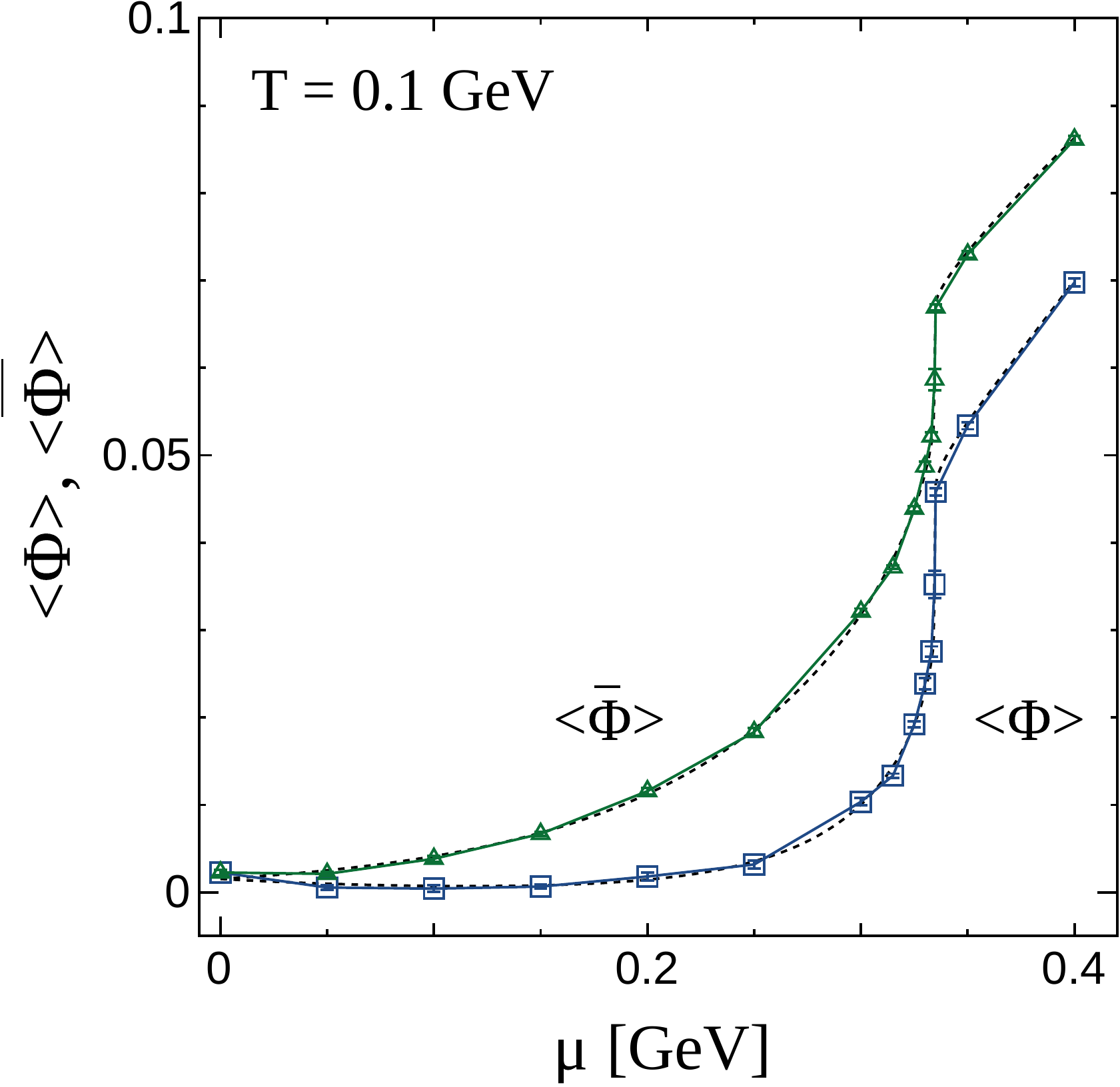}
\caption{The average phase factor (left) and the Polyakov loop and its conjugate
(right) in the PNJL model.}
\label{Fig:PNJL}
\end{figure}

In Fig.~\ref{Fig:PNJL}, we show the results of
path optimization of the PNJL model
at $T=100\,\mathrm{MeV}$ and volume $k=64$~\cite{Kashiwa2018a}.
We take the the diagonal gauge, and $A_3$ and $A_8$ are complexified,
while $\sigma$ and $\vec{\pi}$ are kept to be real.
This choice corresponds to the complexification in the diagonalized gauge
in the 0+1 dimensional QCD, discussed in the previous section.
At around the transition chemical potential, $\mu \simeq 330\,\mathrm{MeV}$,
the average phase factor is suppressed without optimization,
while it becomes almost unity after optimization.
The finite volume results converge to the mean field results
with increasing volume.
The phase transition signal, the rapid increase of the Polyakov loop ($\Phi$)
and its conjugate ($\overline{\Phi}$),
is described well on the optimized path as found in the right panel
of Fig.~\ref{Fig:PNJL}.

\section{Summary}

The sign problem is a grand challenge in theoretical physics,
and we encounter it in many subjects in physics.
Recent development of complexified variable methods
such as the Lefschetz thimble method, complex Langevin method,
and the path optimization method encourages us to tackle the sign problem,
while these methods may be still premature.

In this proceedings, we have discussed the sign problem in field theories
by using the path optimization method with use of the neural network.
In the path optimization method,
we complexify integration variables as $x_i \to z_i=x_i+iy_i$.
The integral path (manifold) is specified by the imaginary part 
as a function of the real parts $y_i(x)$,
and is optimized to evade the sign problem.
We can explicitly parameterize the imaginary part of variables,
$y_i=y_i(\{x\})$, when the number of degrees of freedom is small.
By comparison, it is not practical to prepare and optimize
explicit functions by hand in the case with many variables
such as in field theories.
In such cases, the neural network used in machine learning is useful.
We have adopted a neural network with a single hidden layer.
It should be noted that even with a single hidden layer,
the neural network can describe any functions
in the large unit number limit.
We have demonstrated the usefulness of the path optimization
with use of the neural network in
the complex $\phi^4$ theory at finite $\mu$
the 0+1 dimensional QCD at finite $\mu$,
and the Polyakov loop extended Nambu-Jona-Lasinio model with homogeneous field
ansatz.
In all of these cases,
the average phase factor is enhanced
and the observables are obtained precisely.

One of the problems in the path optimization is the numerical cost.
Since we aim to reduce the sum of complex phases
from the Boltzmann weight $\exp(-S)$ and the measure or Jacobian $J(z)$
simultaneously, we need to calculate the Jacobian,
where the numerical cost is $\mathcal{O}(N^3)$ 
with $N$ being the number of variables.
In order to reduce the cost,
it is helpful to assume the function form of the imaginary part.
For example, the imaginary part is assumed to be a function of the real part
on the same lattice site in the Thirring model in Ref.~\cite{Lawrence},
and it is assumed to be a function of the real part
in the same and the nearest neighbor sites in 0+1 dimensional
$\phi^4$ theory in Ref.~\cite{Bursa2018}.
These assumptions have been demonstrated to work well
in enhancing the average phase factor.
When the Jacobian matrix becomes sparse, we can save the numerical cost.

It is also interesting to apply the deep learning,
where the number of hidden layers is larger than three.
If the integration path is very complex as human brain
with around 7 layers~\cite{Human} cannot imagine,
we may have to invoke deep learning to find the path.


\bigskip

This work is supported in part by the Grants-in-Aid for Scientific Research
from JSPS (Nos. 15H03663, 16K05350, 18K03618),
and by the Yukawa International Program for Quark-hadron Sciences (YIPQS).

\end{document}